\documentclass[12pt,journal,onecolumn ]{IEEEtran}
\usepackage{tikz}
\usetikzlibrary{tikzmark,calc,fit}

 \usepackage{tkz-graph}
\usetikzlibrary{arrows}
\usepackage{tikz}
\usetikzlibrary{positioning, arrows}

\usepackage{url}
\usepackage{enumerate}
 \usepackage{hyperref}

\usepackage[utf8]{inputenc}
\usepackage[margin=1in]{geometry}
\usepackage{graphicx}
\usepackage{amsmath,amssymb,amsthm,authblk,mathrsfs,subfigure}
\usepackage[acronym]{glossaries}
\usepackage{color}
\usepackage[english]{babel}
\usepackage{comment}
\usepackage[nottoc]{tocbibind} 
\usepackage[outdir=./]{epstopdf}

 \usepackage{thmtools}

\usepackage{setspace}

\newcommand{\be}{\begin{equation}}
\newcommand{\ee}{\end{equation}}

\newcommand {\R}{\mathbb R}

\theoremstyle{plain}

\makeatletter
\newcommand{\sq}[1]{\mathbin{\mathpalette\make@circled{#1}}} 
\newcommand{\make@circled}[2]{%
	\ooalign{$\m@th#1\smallbigcirc{#1}$\cr\hidewidth$\m@th#1#2$\hidewidth\cr}%
}
\newcommand{\smallbigcirc}[1]{%
	\vcenter{\hbox{\scalebox{1.2}{$\m@th#1\square$}}}%
}
\makeatother

\declaretheorem[name={Example},qed={\lower-0.3ex\hbox{$\square$}} ] {Example}

\doublespace 

\title{Negative feedback and oscillations in a model for mRNA translation}
\author{Aliza Ehrman, Thomas Kriecherbauer,  Lars Gruene  
  and   Michael Margaliot\thanks{AE and MM (michaelm@tauex.tau.ac.il) are  with the School of Electrical \& Computer Engineering, Tel Aviv University, 69978, Israel. TK and LG are with the
Mathematical Institute, University of Bayreuth, Germany. 
This research   is partially supported by    research grants  from  the~ISF and the DFG (ISF 221/24, GR 1569/24-1,  and KR 1673/7-1).}}
 
\newtheorem{theorem}{Theorem}

\newtheorem{proposition}[theorem]{Proposition}

\newacronym{rfm}{RFM}{Ribosome Flow Model}
\newacronym{pmp}{PMP}{Pontryagin Maximum Principle}
\newacronym{tas}{TASEP}{Totally Asymmetric Simple Exclusion Process}

\definecolor{darkgreen}{rgb}{0, 0.6, 0}

\begin{document}
	\maketitle 

\begin{abstract}

The ribosome flow model (RFM) is a phenomenological model for the unidirectional flow of particles along a~1D chain of~$n$ sites. 
The~RFM has  been extensively used to study the dynamics of ribosome flow along a single mRNA molecule during translation. In this case, the particles model ribosomes and each site corresponds to a consecutive group of codons. Networks of interconnected RFMs have been used to model and analyze large-scale translation in the cell and, in particular, the effects of competition for shared resources.   Here, we analyze the RFM with a negative feedback connection from the protein production rate to the initiation rate. This models, for example, the production of proteins that inhibit the translation of their own~mRNA.
 The RFM with negative feedback is a  2-cooperative  dynamic system, i.e. its flow 
maps the set of vectors with up to one sign variation to itself. 
Using tools from the theory of 2-cooperative dynamical systems, we provide a simple condition guaranteeing that the closed-loop system admits at least one non-trivial periodic solution. When this condition holds, we also explicitly characterize a large set of initial conditions such that any solution emanating from this set converges to a non-trivial periodic solution. Such a solution corresponds to a periodic pattern of ribosome densities along the mRNA, and to a periodic pattern of protein production. 
\end{abstract}

\begin{IEEEkeywords}
 mRNA translation,   
 periodic gene expression, 
 2-cooperative systems. 
\end{IEEEkeywords}



\section{Introduction}

Living cells produce proteins using
RNA molecules as templates. Three consecutive
nucleotides in the mRNA form a codon that encodes for a specific amino-acid. Complex molecular machines, called ribosomes, scan the mRNA
 sequentially, codon by codon, and generate a chain of corresponding amino-acids that eventually becomes the coded protein. Every ribosome that reaches the stop codon releases the chain of amino acids, so the rate of ribosomes leaving the chain is the protein production rate. 
To pipeline this process, several ribosomes may decode the same mRNA molecule simultaneously, 
thus generating a polysome.

The dynamics of ribosome flow along the mRNA is determined by various bio-physical properties including the mRNA sequence, abundance of decoding factors, and the interactions between   ribosomes. 
Indeed, since every ribosome must follow the mRNA in a sequential manner, a slow-moving ribosome may lead to the generation of ``traffic jams'' of ribosomes behind it.  Such traffic jams have been implicated with various diseases including Huntington disease~\cite{Subramaniam2021}, ~\cite{neurojams}.

  A fundamental problem in biology and medicine is understanding the regulation and dynamics of ribosome flow and the protein production rate. 
  Rigorous analysis of ribosome flow
  is also important in scientific
  fields that include natural or artificial manipulations of 
  the translation machinery such as biotechnology,
  synthetic biology,
  and mRNA viruses.   

  \subsection{Modeling ribosome flow}
Computational modeling of ribosome flow is attracting considerable interest~\cite{VONDERHAAR2012e201204002,dykeman_stocahstic_model,zur_survey}.  This is motivated also by new experimental methods, like ribosome profiling~\cite{ingolia_global}, that provide   considerable information  on   which~mRNAs are being actively translated in a cell at a given time.

The totally asymmetric simple exclusion process~(TASEP)~\cite{solvers_guide,kriecherbauer_krug2010,TASEP_book} describes the flow of particles along a 1D lattice with~$n$ sites. Each site may be empty or contain a particle, and particles hop randomly in a unidirectional manner, that is, from site~$i$ to site~$i+1$. Importantly, a particle can only hop into an empty site (simple exclusion), yielding an intricate indirect coupling between the particles. In particular, if a particle is ``stuck'' at a site for an extended period of time, then other particles will accumulate in the sites behind it, thus generating a ``traffic jam''. TASEP has found many useful applications, including modeling the flow of ribosomes along mRNA~\cite{TASEP_tutorial_2011} and other intracellular transport processes~\cite{RevModPhys.85.135}, pedestrian and vehicular traffic~\cite{TASEP_PEDES}, the movement of ants, and numerous other natural and artificial phenomena~\cite{TASEP_book}.  However, rigorous analysis of~TASEP in the case where the transition rates between sites are different is non-trivial. 
Existing results, for example, on contraction in TASEP~\cite{TASEP_RANDOM_ATTRACTION,TASEP_PERIODIC_ATTRACTOR} do not provide the level of detail that deterministic models can provide.

The ribosome flow model~(RFM) is a deterministic model for the flow of ribosomes along the mRNA molecule. It can be derived as a dynamic mean-field approximation of TASEP~\cite{reuveni2011genome}, and is composed of~$n$ non-linear first-order ODEs. The RFM can also be derived via a special finite volume spatial discretization of widely used hyperbolic PDE flow models~\cite{RFM_FROM_PDE}. The RFM can 
be interpreted as a  compartmental  chemical reaction network~(CRN) 
  with transition rates that depend on the amount
of particles and free space in various compartments~\cite{RFM_as_CRN},   
  and can also be represented as a port-Hamiltonian system~\cite{RFM_AS_PORT}. 
 
The~RFM is highly amenable to analysis using tools from  systems and control theory.
The~RFM is a contractive system~\cite{sontag_cotraction_tutorial},   a cooperative system~\cite{margaliot2012stability}, and also a totally positive differential system~\cite{margaliot2019revisiting}.

  The RFM and its variants have been used extensively to model and analyze the flow of ribosomes along the mRNA during translation (see, e.g.,~\cite{rfm_max,rfm_sense,EYAL_RFMD1,alexander2017,rfmr_2015,down_reg_mrna,randon_rfm,Aditi_abortions,Aditi_extended_2022,Ortho_RFM}). In particular, Ref.~\cite{rfm_feedback} studied the RFM with positive feedback from the production rate to the initiation rate, as a model of  ribosome recycling. It was shown that the dynamic behavior of this model is simple: the  closed-loop system  admits a unique equilibrium that is globally asymptotically stable. 

  Interconnected RFM networks have been used to model and analyze large-scale translation in the cell, and in particular the effect of competition for shared resources such as ribosomes and tRNA molecules~\cite{allgower_RFM,Raveh2016,nani,aditi_networks,fierce_compete}. 
 Models based on networks of~RFMs
have also been validated experimentally. It was shown that such a model can predict the density of ribosomes along different mRNAs, the protein levels of different genes, and can even be used to generate ribosomal traffic jams (see, for example, \cite{Zur2020,reuveni2011genome,HALTER2017267}).  

In the case where the entry, exit, and elongation rates in  the RFM are time-varying and jointly periodic, with a common period~$T$,
the dynamics admit a unique $T$-periodic solution that is globally exponentially stable~\cite{RFM_entrain}. Interpreting
the rates as a periodic excitation (e.g., due to the periodic cell-cycle process)
implies that the 
state variables in the RFM, and thus also the protein production rate, entrain
to the periodic excitation (see also~\cite{entrain_trans} for some related considerations).

Gene expression is tightly regulated, and negative feedback loops
are common motifs that can drive the protein production rate to a desired level, as well as
reduce expression noise
and intercellular variability in protein levels~\cite{negative_feedback_noise}.

Gene expression is highly regulated to ensure that the right proteins are made in the right places within the cell.
Here, we are interested in analyzing the effects of negative feedback in the level of translation. 
For example, bacteria use
RNA binding proteins~(RBPs) as regulators of many
cellular processes, including  
translation, and  most~RBPs affect  translation
  at the level of translation initiation. Many   translation-related RBPs    
inhibit their own mRNA~\cite{RNA_BIN_PRO,review_protein_reg}.

 To study this process, 
we analyze for the first time the~RFM with a  negative feedback interconnection
from the protein production rate to the initiation rate.   
We show that:
\begin{enumerate}
    \item The closed-loop system admits a unique equilibrium point~$e$. This corresponds to a   steady-state where the ribosome  flow into and out of each site are equal, and thus to a constant  protein production rate;
    \item Under a simple and easy to verify sufficient condition on the local (in)stability of the system at~$e$, the closed-loop system admits a periodic solution, and we provide an explicit description of a set~$S$ such that any solution emanating from~$S$ converges to a (non-trivial) periodic  solution.
\end{enumerate}
These results suggest that negative feedback in the level of  translation only  can provide a simple cellular  mechanism for producing self-sustained (i.e., unforced) oscillations in protein production.
This agrees with the findings 
in~\cite{ginossar_tran_mito} that used ribosome profiling to conclude that 
gene-specific translational repression controls the mitotic proteome, complementing post-translational mechanisms for inactivating protein function.

The fact that negative feedback may lead to oscillations is not surprising, but rigorously proving this in $n$-dimensional 
non-linear systems like the~RFM is non-trivial. 
Note also that property~2) above implies that the closed-loop system is not contractive  nor cooperative. 
Our analysis is based on showing that the closed-loop  system is a strongly 2-cooperative system~\cite{Eyal_k_posi,KATZ_GIORDANO_MARG_2COOP}.  
  Roughly speaking, strongly 2-cooperative dynamic systems map the set of vectors with up to one sign variation to itself. Such systems have a {Poincar\'{e}-Bendixson} property, which implies that bounded solutions which do not converge to an equilibrium converge to a periodic orbit. 
Ref.~\cite{KATZ_GIORDANO_MARG_2COOP} derived specific conditions on the Jacobian at the equilibrium such that periodic orbits exist, and in this case  also  described  a set of initial conditions that lead to a periodic solution.

 \subsection{Negative feedback and cellular oscillations  }
The emergence of sustained oscillations (via convergence to periodic orbits) in high-dimensional nonlinear dynamical systems is a highly non-trivial question with important applications in systems biology, including the understanding of bio-molecular oscillators ruling cell life-cycle and metabolism, as well as circadian rhythms in hormone secretion, body temperature and metabolic functions~\cite{goldbeter_book}.  

 Many cellular oscillatory processes, including certain circadian oscillators,   are  based on the transcription-translation feedback 
loop~(TTFL), that includes both positive and negative feedback loops~\cite{ttfl_circadian_oscillators}. The recurring motifs of negative feedback loops, positive feedback loops, and reciprocal regulation play a crucial role in timing and regulating   cell-cycle oscillations, and it is known that sufficiently strong  
negative feedback  induces oscillations~\cite{FERRELL2013676}. 
For a very readable account of the role of feedback loops in biological oscillators, see~\cite[Chapter 6]{uri_alon_systems_bio}.

 The p53–MDM2   loop is a well-known negative feedback circuit that plays a crucial role in regulating cell fate decisions.
  p53 activates the transcription of many genes,  including mdm2.    MDM2 inhibits the  transcriptional activity of p53 and    increases
degradation of p53. 
This mechanism yields    oscillations or pulses in p53 levels in response to DNA damage~\cite{osci_p53_pnas2000,lahav_2008,Lahav_2011}.

A negative feedback mechanism in transcription 
induces oscillations in the notch effector Hes1~\cite{hirata_science_2002}.  A ``minimal'' mathematical model for 
this phenomena, that includes two ordinary differential equations and a time delay in the negative feedback loop, was derived in~\cite{Jensen_HES1}.

An important goal in
synthetic biology is building   genetic oscillators that can be used for correctly timing and synchronizing  various processes. 
   Elowitz and    Leibler~\cite{repress_2000}
designed an artificial oscillator in  E. coli 
  consisting of a loop with three transcriptional repressors.  
Negative feedback loops are known to be 
a key design principle  for synthetic biological oscillators~\cite{syn_osci_2020}, but constructing reliable biomolecular oscillators with tunable amplitude and phase is still an open challenge.

 The model that we explore here suggests that oscillations can also be generated by direct negative feedback   from the produced protein to the  mRNA initiation rate.

The remainder of this paper is organized as follows. The next section describes the~RFM with negative feedback. Section~\ref{sec:main} describes the main analysis results. The last section concludes.  The proofs of the main results are placed in the Appendix~1. 

\section{The model}
The dynamic equations of the ribosome flow model~(RFM) with~$n$ sites  are:
\begin{align}\label{eq:simple_rfm}
\dot x_1(t)&= \lambda_0  (1-x_1(t)) -\lambda_1  x_1(t)(1-x_2(t)),\nonumber\\
\dot x_2(t)&= \lambda_1  x_1(t)(1-x_2(t))- \lambda_2(t) x_2(t)(1-x_3(t)),\nonumber\\
&\vdots\\
\dot x_n(t)&= \lambda_{n-1}  x_{n-1}(t)(1-x_n(t))- \lambda_n  x_n(t) ,\nonumber
\end{align}
where~$x_i(t)\in[0,1]$ is  the ribosome density levels  at site $i$ at time~$t$, 
and~$\lambda_i >0, i=0,...,n$ are the translation rates from site~$i$ to 
site~$i+1$ at time~$t$. 
 Note that the~$x_i$s are dimensionless and the~$\lambda_i$s have units of~$1 /   \text{time}$.

To explain this non-linear model, consider the general equation
\be\label{eq:Lgene}
\dot x_i(t) = \lambda_{i-1} x_{i-1}(t) (1-x_i(t)) -\lambda_{i}  x_i(t)(1-x_{i+1}(t)).
\ee
This can be explained as follows. The term~$\lambda_{i-1}  x_{i-1}(1-x_i)$ is the flow rate from site~$i-1$ to site~$i$. This depends on: (1)~the parameter~$\lambda_{i-1}$ that models ``how easy it is'' for a ribosome to move from site~$i-1$ to site~$i$ (e.g., due to the abundance of the cognate     tRNA molecules); (2)~the density~$x_{i-1}$ at site~$i-1$; and
(3) the ``free space''~$1-x_i$ at site~$i$. 
This is a ``soft'' version of the simple exclusion principle: as the density in site~$i$ increases, the flow   into this site decreases. Similarly, the  second term on the right-hand side of~\eqref{eq:Lgene} is the flow rate from site~$i$ to site~$i+1$.

To study large-scale translation in the cell 
using interconnections of RFMs, 
Ref.~\cite{rfm_feedback} 
introduced  the 
  ribosome flow model with input and output~(RFMIO).
  This is derived by adding a scalar  input function
and a scalar  output function
to the~RFM. The input~$u(t) \geq 0$ controls the initiation
rate to the mRNA at time~$t$, and the output~$y(t)$ is the protein production rate at time~$t$. Thus, the RFMIO is described by
\begin{align}\label{eq:rfmio}
\dot x_1(t)&= \lambda_0   (1-x_1(t)) u(t) -\lambda_1 x_1(t)(1-x_2(t)),\nonumber\\
\dot x_2(t)&= \lambda_1 x_1(t)(1-x_2(t))- \lambda_2 x_2(t)(1-x_3(t)),\nonumber\\
&\vdots\\
\dot x_n(t)&= \lambda_{n-1} x_{n-1}(t)(1-x_n(t))- \lambda_n x_n(t) ,\nonumber\\
y(t)&=\lambda_n x_n(t)  . \nonumber
\end{align}

 \subsection*{RFM with negative feedback} 
 
Here, we consider for the first time  the RFMIO~\eqref{eq:rfmio} 
where the input~$u(t)$ is given by
\be\label{eq:nega_feed}
u(t)=k g(y(t)),
\ee
with~$k>0$ and~$g:\R_{\geq 0}\to \R_{>0}$
is a continuously differentiable function 
such that~$g'(z)=\frac{  d g (z) }{ dz} <0$ for any~$z >0 $. Thus,~$g$ is a monotonically decreasing function. We refer to~$k$ as the feedback gain.
This models negative feedback directly from the protein
production rate~$y(t)$ to the RFM initiation rate that is controlled by the input~$u(t)$ (see Fig.~\ref{fig:nega}). 

 Note that~\eqref{eq:nega_feed} captures many negative feedback functions used in models from 
systems biology  and in particular 
gene regulation. For example,
$g(y)=-a y$, with $a>0$ (negative linear feedback), 
$g(y)=e^{-a y} $, with $a>0$ (exponential decay),
 $g(y)= \frac{ 1}{ a+y}$, with~$a >0$ (Michaelis-Menten inhibition), 
 $g(y)=\frac{1}{1+(y/a)^\ell}$, with~$a,\ell>0$ (Hill type inhibition), 
 $g(y)=1-\tanh(ay)$, with~$a>0$, 
 and more.

\begin{figure} 
\begin{center}
\begin{tikzpicture}[
    bottomflat/.style={
        append after command={%
            \pgfextra
                \fill[fill=#1] (\tikzlastnode.south west)   |- (\tikzlastnode.north) -| (\tikzlastnode.east) [sharp corners] |- cycle;
                \draw[rounded corners] (\tikzlastnode.south west) |- (\tikzlastnode.north) -| (\tikzlastnode.south east);
            \endpgfextra}},
    topflat/.style={
        append after command={%
            \pgfextra
                \fill[fill=#1] (\tikzlastnode.north east)  |- (\tikzlastnode.south) -| (\tikzlastnode.west) [sharp corners] |- cycle;
                \draw  (\tikzlastnode.north east) |- (\tikzlastnode.south) -| (\tikzlastnode.north west);
            \endpgfextra}},
    leftflat/.style={
        append after command={%
            \pgfextra
                \fill[fill=#1] (\tikzlastnode.north west) [rounded corners] -| (\tikzlastnode.east) |- (\tikzlastnode.south) [sharp corners] -| cycle;
                \draw[rounded corners] (\tikzlastnode.north west) -| (\tikzlastnode.east) |- (\tikzlastnode.south west);
            \endpgfextra}},
    rightflat/.style={
        append after command={%
            \pgfextra
                \fill[fill=#1] (\tikzlastnode.south east)   -| (\tikzlastnode.west) |- (\tikzlastnode.north) [sharp corners] -| cycle;
                \draw[rounded corners] (\tikzlastnode.south east) -| (\tikzlastnode.west) |- (\tikzlastnode.north east);
            \endpgfextra}},
  block/.style = {draw, minimum width=2cm, minimum height=1.0cm, align=right},
  arrow/.style = {->, thick}
]

\path node[topflat=white!20] (A) {site $1$};
\path node[topflat=white!20, right=20mm of A] (B) {site $2$};
 \path node[topflat=white!20, right=20mm of B] (C) {site $3$};
\path node[topflat=white!20, right=20mm of C] (D) {site $ n$};

\node [right=20mm of D] (E) {};
\node [left=20mm of A] (ZERO) {};

    \path (C) -- node[auto=false]{\ldots} (D);
     \draw[bend left,->]  (ZERO) to node [xshift=0.1cm,yshift=0.5cm] {$\lambda_0(1-x_1)u$} (A);
    \draw[bend left,->]  (A) to node [auto] {$\lambda_1x_1(1-x_2)$} (B);
    \draw[bend left,->]  (B) to node [auto] {$\lambda_2x_2(1-x_3)$} (C);
    \draw[bend left,->]  (D) to node [auto] {$y=\lambda_nx_n$} (E);

      \node[block, below=1cm of B,xshift=1cm] (feed)
{
 $u=kg(y) $  
};
\draw[->](E.west) |- node[above, xshift=-4mm] {} (feed.east);
\draw[->](feed.west) -| node[above, xshift=-4mm] {} (ZERO.west);
\end{tikzpicture}
\caption{ RFMIO with   feedback~$u=kg(y)$.} \label{fig:nega}
\end{center}
\end{figure}

The closed-loop system that 
arises from~\eqref{eq:rfmio} by eliminating~$y$ and substituting~$u$ from~\eqref{eq:nega_feed}
is 
\begin{align}\label{eq:closedloop}
\dot x_1(t)&= \lambda_0 k g(\lambda_nx_n(t)) (1-x_1(t)) -\lambda_1 x_1(t)(1-x_2(t)),\nonumber\\
\dot x_2(t)&= \lambda_1 x_1(t)(1-x_2(t))- \lambda_2 x_2(t)(1-x_3(t)),\nonumber\\
&\vdots\\
\dot x_n(t)&= \lambda_{n-1} x_{n-1}(t)(1-x_n(t))- \lambda_n x_n(t) \nonumber.
\end{align}

Let~$f:[0,1]^n \to\R^n$ be the vector field on the right-hand side of~\eqref{eq:closedloop}, so that we can write the closed-loop system as
\[
\dot x=f(x),
\]
and let
\[
J(x):=\frac{\partial}{\partial x}f(x)
\]
be the Jacobian 
of the vector field~$f$.  For an initial condition~$x(0)=a$, let~$x(t,a)$ denote the solution of the closed-loop system at time~$t\geq0$.

\section{Main results}\label{sec:main}

We can now state  our main results. The proofs of these results are given in   Appendix~1. 
Our first result guarantees in particular  that the unit cube~$[0,1]^n$ is a forward  invariant set for the closed-loop system. In other words, if we initialize the state-variables~$x_i(0)$ at time zero with admissible density values, that is, all values are in the range~$[0,1]$ 
then they maintain admissible values for all~$t>0$. 

\begin{proposition}\label{prop:unitcube}
For any initial  condition~$a\in[0,1]^n$, we have that~$x(t,a)\in(0,1)^n$ for all~$t>0$.
\end{proposition}

Our second 
 main result  shows that the closed-loop system admits a unique equilibrium~$e\in[0,1]^n$. At this equilibrium, that is, when~$x_i(t)=e_i$ for all~$i$,  the flow into each site 
is equal to the flow out of the site,  and thus the density at each site remains unchanged. Furthermore, for any feedback gain~$k>0$ sufficiently small   the  equilibrium is globally asymptotically stable, that is, the density along the mRNA  will converge  to~$e$ from  any initial condition. 
\begin{proposition}\label{prop:uniequ_e}
   The closed-loop system described by~\eqref{eq:rfmio} and~\eqref{eq:nega_feed}  admits a unique equilibrium~$e$ in the unit cube, and~$e\in(0,1)^n$. If the feedback gain~$k>0$ is sufficiently  small then  any solution of the closed-loop system converges to~$e$. 
\end{proposition}

Our third
 main result provides a simple 
condition guaranteeing that the closed-loop system admits a (non-trivial)  periodic orbit. To state it, we recall the notion of the weak number of sign variations in a vector~\cite{Fallat2011TNBook}.

If~$z\in\R^n$ has no zero entries then the number of sign variation in~$z$, denoted~$\sigma(z)$,  is the number of indexes~$i$ such that~$z_iz_{i+1}<0$.  For example, for~$z=\begin{bmatrix}
    1&-2&4
\end{bmatrix}^\top$, we have~$z_1z_2<0$ and~$z_2z_3<0$, so~$\sigma(z)=2$.  
More generally, for any vector~$z\in\R^n$, with~$z\not =0$, the   weak number of sign variations in~$z$ 
is defined as~$s^-(z):=\sigma(\tilde z)$, where~$\tilde z$ is the vector obtained from~$z$ after deleting all its zero entries. For example, for~$z=\begin{bmatrix}
  0&  1&-2&0&4
\end{bmatrix}^\top$, we have~$\tilde z=\begin{bmatrix}
    1&-2& &4
\end{bmatrix}^\top$, so~$s^-(z)=\sigma(\tilde z)=2$. For the zero vector,   define~$s^-(0):=0$. Note that for any~$z\in\R^n$ and any non-zero scalar~$c$, we have~$s^-(cz)=s^-(z)$. 

An eigenvalue~$\lambda$ of a matrix~$A\in\R^{n\times n} $ is called unstable if the real part of~$\lambda$ is positive. 
\begin{theorem}\label{thm:main}
Let~$e\in ( 0,1 ) ^n$ denote the unique equilibrium of the closed-loop system described by~\eqref{eq:rfmio} and~\eqref{eq:nega_feed}.
   If~$J(e)=\frac{\partial}{\partial x}f(e)$ has at least two unstable eigenvalues
    then the following property holds. For any initial condition~$x(0)\in[0,1]^n$ such that~$x(0) \not = e$
    and~$s^-(x(0)-e)\leq 1$ the corresponding solution~$x(t,x(0))$ converges to a non-trivial periodic orbit. 
\end{theorem}

Note that since~$e\in(0,1)^n$, we have~$s^-(0-e)=0$ implying that 
 if~$J(e) $ has at least two unstable eigenvalues then the solution~$x(t,0)$ 
 always converges to a non-trivial periodic orbit. The zero initial condition corresponds in our model to zero  ribosomes along the mRNA molecule at the initial time, which is the case for a newly synthesized mRNA. 
 
The next example demonstrates Theorem~\ref{thm:main} using a numerical simulation. The Matlab code for generating this simulation is provided in Appendix~2.

\begin{figure}
    \centering
    \includegraphics[scale=0.85]{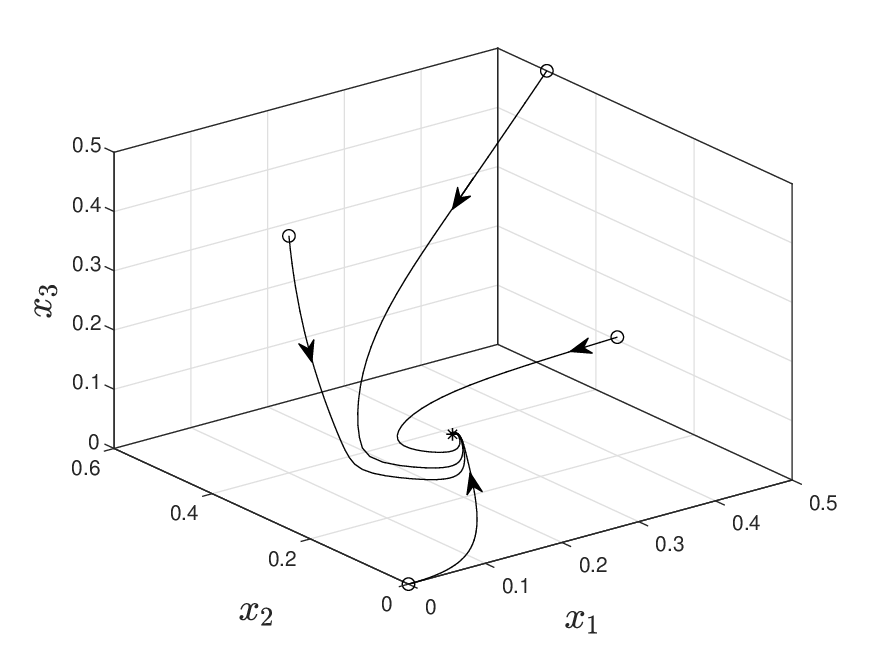}
    \caption{  Convergence of several trajectories of the closed-loop system in Example~\ref{exa:neg3} to the equilibrium point with~$k=0.1$. The initial conditions are marked with~$\circ$ and the   equilibrium point~$e$ is marked by~$*$.}
    \label{fig:simu3ksmall}
\end{figure}
\begin{figure}
    \centering
    \includegraphics[scale=0.85]{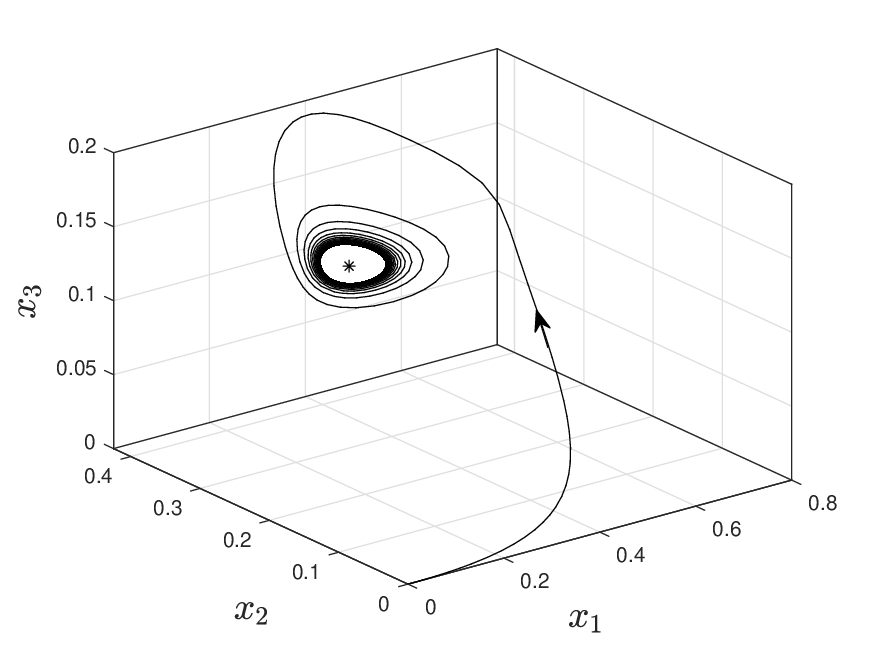}
    \caption{ Convergence of a trajectory of the closed-loop system in Example~\ref{exa:neg3} to a periodic orbit with~$k=1$.} The equilibrium point~$e$ is marked by~$*$.
    \label{fig:simu3}
\end{figure}

\begin{Example}\label{exa:neg3}
    Consider the closed-loop system with~$n=3$, 
$\lambda_0=12$, $\lambda_1=6$, $\lambda_2=5$,
$\lambda_3=8$, and~$u(t)=kg(y(t))$,   where~$g(s)=\frac1{ 1+s^{20}}$. 
When~$k=0.1$, 
 a calculation shows that
$
e=\begin{bmatrix}
    0.1855    &0.2004    &0.1113
\end{bmatrix}^\top
$ (all numerical values in this paper 
are to 4-digit accuracy),
and the eigenvalues of $J(e)$ are 
\[ -14.9362 ,\;
-2.7564 + 6.0039i,\;
  -2.7564 - 6.0039i.
\]
Note that these all have negative real parts.
Fig.~\ref{fig:simu3ksmall}
depicts several trajectories of the system, and it may be seen that they all converge to~$e$. 

 When~$k=1$, 
 a calculation shows that
$
e=\begin{bmatrix}
     0.2472  &  0.2560 &   0.1380
\end{bmatrix}^\top,
$
 and the eigenvalues of~$J(e)$ are
\[
 -21.0188 , \; 0.0075 + 11.9317i,\; 0.0075 - 11.9317i,
\]
so~$J(e) $ admits two unstable eigenvalues. 
Fig.~\ref{fig:simu3} depicts the solution~$x(t,a)$  of the closed-loop system emanating from the initial condition~$a=0$. Note that~$s^-(a-e)=0$.
It may be seen that this solution does not converge to~$e$,
but rather converges to a periodic orbit. 
 \end{Example}

 Note that Theorem~\ref{thm:main} holds for the closed-loop system  with any dimension~$n$. In Example~\ref{exa:neg3} we considered the special case~$n=3$ only because
 in this case it is easy to plot the solutions of the system.

 Our theoretical results can be easily applied to RFMs with transition rates
that 
are based on biological data. The analysis flow in such a case is demonstrated in the following example. The Matlab code for the following calculations  is provided in Appendix~3.

\begin{figure}
    \centering
    \includegraphics[scale=0.85]{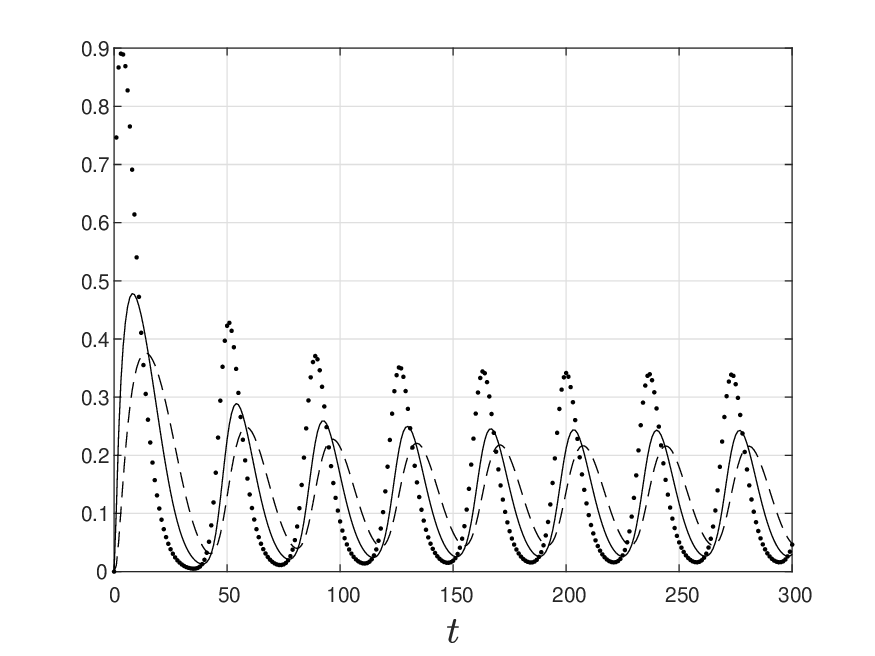}
    \caption{ The state-variables~$x_1(t)$ (dotted),
    $x_2(t)$ (dashed), and~$x_3(t)$ (solid)  as a function of~$t$ in Example~\ref{exa:bio}. Here~$t$ is in sec. To avoid cluttering,~$x_4(t)$ and~$x_5(t)$ are not plotted.
    \label{fig:simu5}}
\end{figure}

\begin{Example}\label{exa:bio}
 
 Consider
 the  RFM derived in~\cite{down_reg_mrna} for the 
S. cerevisiae gene YBL025W that encodes the protein
RRN10, which is related to regulation of RNA polymerase I.
This gene has 145 codons (excluding the stop codon), and the corresponding  
RFM is based on 
dividing the mRNA into 6 consecutive pieces that include 
24, 25, 25, 25, 25,  and   21
  non-overlapping
codons, respectively. 
The first piece corresponds to the initiation stage. 
The corresponding RFM thus includes $n = 5$ sites 
and six rates~$\lambda_0,\dots,\lambda_5$. These rates were estimated using  ribo-seq
data 
(see~\cite{down_reg_mrna}) yielding
\be\label{eq:lam_vals}
\begin{bmatrix}
    \lambda_0&\dots&\lambda_5
\end{bmatrix}=
\begin{bmatrix}
  0.1678
  &0.2572
  &0.2758 
  &0.2514
  &0.2612
  &0.3002
\end{bmatrix},
\ee
with units of~$\text{sec}^{-1}$.
Assume that we close the loop around the corresponding RFMIO with the    negative feedback function 
\[
 g(y)= 1-\tanh(a y ),
\]
with~$ a>0$. This function takes values in~$[0,1]$, and
for large  values of~$a$ it decays    quickly with~$y$, resembling an on-off switch. 
Note that
$
g'(y)=-\frac{a}{\cosh^2(ay)}
$.

The  equilibrium point~$e $ satisfies the equations 
\begin{align*}  
 \lambda_0 k(1-\tanh(a \lambda_5 e_5))  (1-e_1)   &=\lambda_1 e_1(1-e_2)\nonumber\\
 &=  \lambda_2 e_2(1-e_3)\nonumber \\
&= \lambda_3 e_3(1-e_4)\\
 &= \lambda_{4} e_{4}(1-e_5)\nonumber\\
 &= \lambda_5 e_5, 
\end{align*}
with the~$\lambda_i$s in~\eqref{eq:lam_vals}. This set of equations can be solved numerically for any~$k,a$, and Proposition~\ref{prop:uniequ_e}
guarantees the existence of a unique solution~$e\in(0,1)^5$. For example, for~$k=10$ and~$a=100$ the unique  solution is
\be\label{eq:sole}
e=\begin{bmatrix}
     0.1038&
    0.0978&
    0.1065&
    0.1003&
    0.0803
\end{bmatrix}^\top.
\ee
 
The Jacobian of the closed-loop system
at $e$ is  
\[
 J(e)=  \left [ \begin{smallmatrix}
-\lambda_0  k g(\lambda_5 e_5) -\lambda_1(1-e_2)& \lambda_1 e_1 &0&0&   \lambda_0\lambda_5 (1-e_1)kg'(\lambda_5 e_5) \\
\lambda_1(1-e_2) & -\lambda_1 e_1 -\lambda_2(1-e_3) & \lambda_2 e_2 &0 &    0   \\
0&\lambda_2(1-e_3) & -\lambda_2 e_2 -\lambda_3(1-e_4) & \lambda_3 e_3 &0 \\
0&0&\lambda_3(1-e_4) & -\lambda_3 e_3 -\lambda_4(1-e_5) & \lambda_4 e_4  \\
0&0&0&  \lambda_{4} (1-e_5)&
-\lambda_{4}e_{4}-\lambda_5
\end{smallmatrix}
\right]. \]
Substituting the numerical values above  and calculating the eigenvalues of~$J(e)$ gives
\[
 0.0112 + 0.1894i,\;
   0.0112 - 0.1894i,\;
  -0.3855 + 0.3072i,\;
  -0.3855 - 0.3072i,\;
  -0.6301  .
\]
 Thus, two eigenvalues have a positive real part and our theoretical results imply that 
 for any initial condition~$x(0)\in[0,1]^5$ such that~$x(0) \not = e$
    and~$s^-(x(0)-e)\leq 1$ the corresponding solution~$x(t,x(0))$ converges to a non-trivial periodic orbit. 
Fig.~\ref{fig:simu5} depicts the solution~$x(t,x(0))$  of the closed-loop system emanating from the initial condition~$x(0)=0$. Note that~$s^-(x(0)-e)=0$.
It may be seen that this solution does not converge to~$e$,
but rather converges to a periodic orbit. Note that after convergence to the periodic pattern, it is straightforward to obtain from the numerical simulation features such as the amplitude and frequency of each density and of the protein production rate. 
\end{Example}

 \section{Discussion}
We considered a mathematical model for mRNA translation with a negative feedback connection
from the protein production rate to the initiation rate. This may model for example  
the effect of  RBPs that inhibit their 
own mRNA, as well as other  gene expression regulation mechanisms like the production  of  proteins that prevent  the assembly of the translation initiation complex. More generally, negative regulation of mRNA translation can be used to fine tune protein synthesis, maintaining cellular homeostasis and adaptivity to
 environmental changes or internal signals.

 Our results show that the closed-loop system admits a unique steady-state~$e$, but if this steady state is locally unstable (more specifically, if the Jacobian of the vector field at~$e$ admits at least two unstable eigenvalues) then non-trivial oscillations appear. These correspond to a periodic pattern of protein production. This may also generate an internal clock 
that can regulate timing in various cellular
processes.

It is interesting to note that Castelo-Szekely et al.~\cite{castelo2017}
have found that tissue specificity in rhythmic gene expression extends to the translatome and contributes to define the identities, the phases and the expression levels of rhythmic protein biosynthesis. Also, it is known that translation oscillates during the periodic cell-cycle program (see, e.g.,~\cite{ginossar_tran_mito}), and our results suggest that this may be generated by a simple negative feedback mechanism.

 More work is needed in order to analyze  the properties of the oscillatory solutions in our model e.g. their amplitude and frequency, and on how these features  depend on the various parameters of the system. 

 Our analysis is based on the fact 
that the RFM with negative feedback is a  $2$-cooperative system. It is natural to ask if this structure is preserved when we add more bio-physical   features into the RFM. We describe two 
possible generalizations that preserve the $2$-cooperative structure. First, 
adding a decay term to each density equation in the RFM, that is, changing~\eqref{eq:Lgene} to 
\[ 
\dot x_i(t) =-c_i x_i(t) + \lambda_{i-1} x_{i-1}(t) (1-x_i(t)) -\lambda_{i}  x_i(t)(1-x_{i+1}(t)) , 
\]
with~$c_i\in\R$.
Indeed, such a modification changes the diagonal terms of the Jacobian of the system, but the diagonal terms play no role in determining $2$-cooperativity 
(see Eq.~\eqref{eq:2coop_matrix} below). 
Of course, the $c_i$s may change the location of the equilibrium~$e$, as well as the spectral properties of~$J(e)$.

 Another and perhaps more interesting 
possibility that preserves $2$-cooperativity is the feedback connection of two RFMIOs, as depicted in Fig.~\ref{fig:rfm_2_inter}, with~$g_1$ a monotonically decreasing function, 
and~$g_2$ a monotonically increasing function.
Such a closed-loop network may be used as 
a model for the interaction between two mRNAs, via the proteins that they produce, and our results allow to determine when oscillations appear in such a
network.

\begin{figure} 
\begin{center}
    \begin{tikzpicture}[
  block/.style = {draw, minimum width=2cm, minimum height=1 cm, align=center},
  arrow/.style = {->, thick}
]
\node[block] (block1) 
{
 $\begin{aligned} 
 \dot x_1  &= \lambda_0(1-x_1)u_1-\lambda_1x_1(1-x_2) \\
 \dot x_2  &= \lambda_1x_1(1-x_2)-\lambda_2x_2(1-x_3) \\
 &\vdots\\
 \dot  x_n   & =  \lambda_{n-1}x_{n-1}(1-x_n)-
 \lambda_nx_n
 \end{aligned} $  
};

 \node[  right=3cm of block1] (blockstam)
{ 
};
\node[  right=1.75cm of block1] (middle_y)
{ 
};


 \node[block, below=2cm of block1](block2)
{
  $\begin{aligned} 
 \dot z_1  &= \eta_0(1-z_1)u_2-\eta_1z_1(1-z_2) \\
 \dot z_2  &= \eta_1z_1(1-z_2)-\eta_2z_2(1-z_3) \\
 &\vdots\\
 \dot  z_m   & =  \eta_{m-1}z_{m-1}(1-z_m)-
 \eta_mz_m
 \end{aligned} $    
};

 \node[block, below right =0.8cm and 0.7cm of  block1 ] (feed_func_g2)
{
  $  
 u_2=g_2(y_1)
   $    
};
 \node[block, below left =0.8cm and 0.7cm of  block1 ] (feed_func_g1)
{
  $  
 u_1=g_1(y_2)
   $    
};

\node[ left=1.75cm of block1] (left1)
{ 
};

\node[  left=1cm of block2] (middle_w)
{ 
};

\draw[arrow] (block1.east) -- node[above,xshift= 2mm] {$y_1=\lambda_nx_n$} (middle_y.center);

\draw[arrow] (left1.center) -- node[above,xshift=-2mm] {$u_1$} (block1.west);
 
\draw[->](middle_y.center) -- node[above, xshift=0mm] {} (feed_func_g2.north);

\draw[->] (block2.west)  -| node[below, xshift=5mm] {$y_2=\eta_m z_m$}   (feed_func_g1.south);

\draw[->] (feed_func_g2.south)  |- node[below, xshift=5mm] { } (block2.east)   ;

\draw[->] (feed_func_g1.north)  -- node[below, yshift=0mm] {} (left1.center);

\end{tikzpicture}\caption{ Feedback interconnection of
two RFMIOs. One RFMIO has dimension~$n$ and rates~$\lambda_i$, and the second has dimension~$m$ and rates~$\eta_i$. They are interconnected via two feedback functions~$g_1$ and~$g_2$. }\label{fig:rfm_2_inter}
\end{center}
\end{figure}

\subsection*{Acknowledgments}
We thank Alexander Ovseevich and Tamir Tuller
 for helpful comments.  
 We are grateful
to the anonymous reviewers and the editor 
for many helpful comments that helped us to improve this  paper. 

\section*{Appendix 1: Proofs}

\emph{Proof of Proposition~\ref{prop:unitcube}.}
Consider the
closed-loop system~\eqref{eq:closedloop}.
Fix an initial condition~$x(0)=a\in[0,1]^n$. Suppose that there exist~$i
\in\{1,\dots,n\}$ and~$t\geq 0$ such that~$x_i(t,a)=0$. 
Then there exists a minimal index~$j$ such that~$x_j(t,a)=0$.
If~$j=1$ then
\[
\dot x_1(t)=\lambda_{0} k  g(\lambda_nx_n(t))>0.
\]
If~$j\in\{2,\dots,n-1\}$ then
\[
\dot x_j(t)=\lambda_{j-1} x_{j-1}(t)>0.
\]
This implies that~$x_\ell(t)>0$ for all~$\ell\in\{1,\dots,n\}$ and all~$t>0$. A similar argument shows that~$x_\ell(t)<1$ for all~$\ell\in\{1,\dots,n\}$ and all~$t>0$, and this completes the proof.

\emph{Proof of Proposition~\ref{prop:uniequ_e}.}
Since~$[0,1]^n$ is convex  and compact, it follows from Proposition~\ref{prop:unitcube} and  Brouwer's fixed point theorem~\cite{kellogg1976constructive} that the closed-loop system admits at least one equilibrium~$e\in[0,1]^n$. Using Proposition~\ref{prop:unitcube} implies that in fact~$e\in(0,1)^n$. 
By~\eqref{eq:closedloop}, the equilibrium satisfies
\begin{align} \label{eq:eq_point}
 \lambda_0  k g(\lambda_n e_n) (1-e_1)   &=\lambda_1 e_1(1-e_2)\nonumber\\
 &=  \lambda_2 e_2(1-e_3)\nonumber \\
&\vdots\\
 &= \lambda_{n-2} e_{n-2}(1-e_{n-1})\nonumber\\
 &= \lambda_{n-1} e_{n-1}(1-e_n)\nonumber\\
 &= \lambda_n e_n.  \nonumber
\end{align}
These equations imply in particular that~$e_n$ uniquely determines~$e_{i}$ for all~$i$. In order to prove uniqueness of the equilibrium, seeking a contradiction, suppose that~$p,q\in(0,1)^n$ are two different equilibrium points. Then~$p_n\not=q_n$, and we may assume that~$p_n>q_n$. By~\eqref{eq:eq_point},
$
e_{n-1}=\frac{\lambda_{n}e_n} {\lambda_{n-1}(1-e_n)} 
$, so~$p_{n-1}>q_{n-1}$. 
Eq.~\eqref{eq:eq_point} also gives
$e_{n-2}= \frac{\lambda_n e_n}{ \lambda_{n-2} (1-e_{n-1})} $,  so~$p_{n-2}>q_{n-2}$. Continuing in this way, we find that~$p_i>q_i$ for all~$i$. Now~\eqref{eq:eq_point} and the fact that~$g$ is decreasing give
\begin{align*}
    \lambda_n p_n &= \lambda_0 k  g(\lambda_n p_n)(1-p_1)\\
  &<  \lambda_0  k g(\lambda_n q_n)(1-q_1)\\
  &= \lambda_n q_n,
\end{align*}
so~$p_n<q_n$. This contradiction shows that the equilibrium is unique.

The Jacobian of the closed-loop system is
\be\label{eq:jacob}
J(x)=\left[ \begin{smallmatrix}
-\lambda_0  k g(\lambda_n x_n) -\lambda_1(1-x_2)& \lambda_1 x_1 &0&0&\dots&0&   \lambda_0\lambda_n (1-x_1)kg'(\lambda_n x_n) \\
\lambda_1(1-x_2) & -\lambda_1 x_1 -\lambda_2(1-x_3) & \lambda_2 x_2 &0 &    \dots&0&0  \\
&&\vdots\\
                0&0&0&0&\dots& \lambda_{n-1} (1-x_n)&-\lambda_{n-1}x_{n-1}-\lambda_n
\end{smallmatrix}\right ] .
\ee
 Recall that~$g'(z) <0$ for any~$z >0 $, so   entry~$(1,n)$ in $J(x)$ is non-positive.
Let~$s_i(x)$ denote the sum of the entries in column~$i$ of~$J(x)$, with the off-diagonal terms taken with absolute value. Then 
\begin{align*}
    s_1(x)  & = -\lambda_0 k  g(\lambda_n x_n) \leq -\lambda_0 k  g(\lambda_n  ),\\
    s_2(x) & =\dots=s_{n-1}(x)=0,\\
    s_n(x) &= -\lambda_n+\lambda_0\lambda_n (1-x_1)k|   g'(\lambda_n x_n)  | \leq
   -\lambda_n - \lambda_0\lambda_n  k   \min_{r\in[0,1]} g'(\lambda_n r )  .
\end{align*}
For any~$k>0$ sufficiently small, we have that~$s_1(x),s_n(x)<0$ for all~$x\in[0,1]^n$. 
Recall that the matrix measure 
induced by the~$L_1$ norm is~$\mu_1(J(x))=\max  \{ s_1(x),\dots,s_n(x) \} $ (see, e.g.,~\cite{sontag_cotraction_tutorial}). Using a scaled~$L_1$ norm, as in~\cite{RFM_entrain}, yields that the system is contractive. This implies that~$e$ is globally asymptotically stable.

\emph{Proof of Theorem~\ref{thm:main}.}
 A 2-cooperative dynamic system has a Jacobian with the following sign pattern 
\be\label{eq:2coop_matrix}
\bar{A}_2:=\left[ \begin{smallmatrix}
* & \geq0 &0&\dots&0&   \leq0 \\
\geq0 & * & \geq0  &    \dots&0&0  \\
0 &\geq0 & *  &    \dots&0&0  \\
&&\vdots\\
0 &0 &0 &    \dots&*&\geq0\\
\leq0 &0 &0 &    \dots&\geq0&*\\
\end{smallmatrix}\right ] 
\ee
where "*" means that the sign of the  entry  is not important.
Since~$g'(z)<0$ for all~$z\geq 0$, Eq.~\eqref{eq:jacob} implies that  
$J(x)$ has the sign pattern of a 2-cooperative system~\cite{Eyal_k_posi} for all~$x\in[0,1]^n$,
and it is irreducible for all~$x\in(0,1)^n$. Using Theorem~2 in~\cite{KATZ_GIORDANO_MARG_2COOP} 
completes the proof of Theorem~\ref{thm:main}.

\section*{Appendix 2: Matlab code for generating Example~\ref{exa:neg3}}
\begin{verbatim}
%%%%% file negtive_RFM

global k  % controller gain 
global lam0 lam1 lam2 lam3 % transition rates along the chain
lam0=12   ;lam1 =6 ; lam2=5 ; lam3= 8  ;

set(gcf,'renderer','painters')
figure (1); hold off;
k=0.1; % low gain
init=[0 0 0]';plot3(0,0,0,'ok');hold on;
[t,x]=ode45(@dyn_negative_RFM,[0 100],init);
plot3(x(:,1),x(:,2),x(:,3),'k');
init=[1/2 1/2 1/2]'; plot3(1/2,1/2,1/2,'ok');hold on;
[t,x]=ode45(@dyn_negative_RFM,[0 100],init);
plot3(x(:,1),x(:,2),x(:,3),'k');

init=[.1 .4 .4]';plot3(.1,.4,.4,'ok');hold on;
[t,x]=ode45(@dyn_negative_RFM,[0 100],init);
plot3(x(:,1),x(:,2),x(:,3),'k');

init=[.4 .2 .2]';plot3(.4,.2,.2,'ok');hold on;
[t,x]=ode45(@dyn_negative_RFM,[0 100],init);
plot3(x(:,1),x(:,2),x(:,3),'k');

box on;
grid on;
plot3(0.1855, 0.2004, 0.1113,'*k'); % plot the eq
 xlabel('$x_1$',Interpreter='latex',FontSize=18);
 ylabel('$x_2$',Interpreter='latex',FontSize=18);
 zlabel('$x_3$',Interpreter='latex',FontSize=18);
%%%%%%%%%%%%%%%%%%%%%%
set(gcf,'renderer','painters')
figure(2); hold off;
k=1;   %high gain 
init=[0 0 0]';
[t,x]=ode45(@dyn_negative_RFM,[0 100],init);
plot3(x(:,1),x(:,2),x(:,3),'k');hold on;
plot3(0.2472 ,0.2560 ,0.1380 ,'*k'); % plot the eq
box on
grid on
xlabel('$x_1$',Interpreter='latex',FontSize=18);  
ylabel('$x_2$',Interpreter='latex',FontSize=18);
zlabel('$x_3$',Interpreter='latex',FontSize=18);
%%%%%%%%%%%%%%%%%%%%%%
function ret=dyn_negative_RFM(t,x)
global k   % controller gain 
global lam0 lam1 lam2 lam3 % transition rates along the chain
x1=x(1);x2=x(2);x3=x(3);
u= k*1/(1+(lam3*x3)^20 ); % the negative feedback 
x1d = lam0*u*(1-x1)-lam1*x1*(1-x2);
x2d = lam1*x1*(1-x2)-lam2*x2*(1-x3);
x3d = lam2*x2*(1-x3)-lam3*x3;
ret=[x1d;x2d;x3d];
%%%%%%%%%%%%%%%%%%%%%%
\end{verbatim}

\section*{Appendix 3: Matlab code for calculating the Jacobian in Example~\ref{exa:bio}}
\begin{verbatim}
 %%%%%%%%%%%%%%%%%
lam0=0.1678;
lam1=0.2572;
lam2=0.2758;
lam3=0.2514;
lam4=0.2612;
lam5=0.3002;
k=10;  
a=100;

%% The eq equations 
%%%%%%%%%%%%%%%%%%%
fun = @(x) [lam0*k* (1-tanh(a*lam5*x(5)))*(1-x(1))-lam1*x(1)*(1-x(2));
lam1*x(1)*(1-x(2))-lam2*x(2)*(1-x(3));
lam2*x(2)*(1-x(3))-lam3*x(3)*(1-x(4));
lam3*x(3)*(1-x(4))-lam4*x(4)*(1-x(5));
lam4*x(4)*(1-x(5))-lam5*x(5) ];
 
x0 = [1/2; 1/2; 1/2; 1/2; 1/2];% an initial guess
[x, fval, exitflag, output] = fsolve(fun, x0,options);
 
e=x; %%%   eq point

% calculate the Jacobain J(e) 
%%%%%%%%%%%%%%%%%%%%%%% 
y=lam5*e(5);
g=1-tanh(a*y) ;
gdot= -a/ (cosh(a*y))^2 ;

J=[-lam0*k*g-lam1*(1-e(2))  lam1*e(1)  0  0  lam0*lam5*(1-e(1))*k*gdot ; 
   lam1*(1-e(2))  -lam1*e(1)-lam2*(1-e(3)) lam2*e(2)   0             0; 
    0           lam2*(1-e(3))   -lam2*e(2)-lam3*(1-e(4)) lam3*e(3)   0;  
    0     0      lam3*(1-e(4))     -lam3*e(3)-lam4*(1-e(5)) lam4*e(4) ;
    0     0             0           lam4*(1-e(5)) -lam4*e(4)-lam5    ];
eig(J)
\end{verbatim}

 \bibliographystyle{IEEEtranS}

\end{document}